# Electron-phonon associated carrier mobility in MgSe and MgTe

Maitry Joshi[1], Trupti K Gajaria[1,2] and Prafulla K. Jha[1,*]

[1]Department of Physics, Faculty of Science, The Maharaja Sayajirao University of Baroda, Vadodara – 390002, Gujarat, India.

[2]School of Science, GSFC University, P. O. Fertilizer Nagar, Vigyan Bhavan, Vadodara – 391750, Gujarat, India.

*prafullaj@yahoo.com

## Abstract

Electron-phonon (E-p) coupling incorporated density functional theory (DFT) based investigation of structural, electronic and vibrational properties of bulk MgSe and MgTe is presented. Electron-phonon coupling is incorporated to understand its effect on charge carrier dynamics. It is observed that the MgTe possesses room temperature hole and electron mobility of 18.7 cm$^2$/Vs and 335 cm$^2$/Vs, respectively; in contrast to this, the bulk MgSe follows reverse trend in temperature dependent carrier mobilities owing to its different scattering rate and electron-phonon coupling profiles. The key feature of the study was to showcase the importance of electron-phonon coupling in determining the carrier mobility and the relative dynamics in the material. Further, the incorporation of e-p coupling softens the electronic and phonon dispersions which is subjected to the inclusion of the interaction between the electrons and the phonons of the systems. The overall results indicate that the incorporation of the electron-phonon coupling is crucial in determining the carrier dynamics of the system which is in excellent agreement with the experimental findings. Both materials possess moderate magnitudes of mobilities that are subjected to the modification by means of changing the dimensional confinement and/or chemical composition that can strongly influence the carrier dynamics by means of altered edge states and coupling parameters.

## 1. INTRODUCTION

The field of materials science is vast and widespread across all cutting-edge technologies and hence, the prediction of different materials properties prior to their synthesis plays a crucial role in saving resources [1]. Out of all materials categories, the semiconductors have won the race owing to their versatile properties. Nasdaq, in its one of the reports has quoted that one of the global tycoons, i.e., the AMD expects market growth of around US$400 billion by 2027 [2]. Carrier transport is the critical and dominating parameter that can be tuned and tailored as per the application specific device requirements. The applicability of the artificial intelligence-based systems significantly depends on the efficiency of the material framework used, thereby promoting significant contributions in the research and development in the allied areas [3]. Extensive research has been carried out globally for finer and clear description of carrier dynamics within the material, as most of the material properties are governed under the umbrella of a central question of interest: how do the carriers behave within a material? [4–6], as it is a game changer in the nanoelectronics, optoelectronics and spintronics based device applications. This points out at the necessity of developing and designing novel materials that possess unique characteristics like rapid on-off states [7], moderate carrier recombination rates [7], high exciton binding energy [8], etc.; all of these parameters are directly dependent on the carrier mobility which in turn has dependence on the electronic band dispersion and the material's mechanical properties.

The group of materials made from metal chalcogenides have been proven to give benchmarking performance under bulk as well as confined dimensions [9–13].

However, the combination of magnesium and chalcogenides have not been studied in-depth for unravelling the carrier dynamics, and, rather the reports are limited to the general ground state properties of the compound [14–21], and very scarce theoretical reporting is available describing carrier dynamics within bulk magnesium chalcogenides. Dillenz et al. [22] have reported significant enhancement in the carrier mobility on insertion of Mg ions in the host material. Several other materials are also reported to show similar results on combining Mg based materials with the host material and enhance thermoelectric efficiency of the system [11,21,23,24]. Furthermore, the thermal conductivity of the MgTe is reported to be significantly lower as compared its other sibling members of group II-VI [25]. Therefore, it is foreseen that the unique band dispersion and phonon dynamics can be a key role player for the carrier dynamics within the material, and unravelling the carrier dynamics within these materials is crucial parameter for developing and deploying high performance energy efficient future devices.

Density functional theory (DFT) [26] has been utilized extensively across the globe for understanding diverse mechanism of materials' underlying properties and governing factors of the same. Theoretically, the carrier dynamics within the materials can be predicted accurately only by addressing the coupling between different energy carriers governing the same. Recent development in the DFT has incorporated one of such crucial coupling parameters known as electron-phonon (e-p) coupling [27]. It has been reported that after inclusion of e-p coupling, the predicted carrier dynamics of the materials gives excellent agreement with the experimentally reported data [28,29]. The e-p interaction plays a crucial role in the determining the electron and lattice dynamics related properties of a system. For example, phenomena such as the resistivity [30] and conventional superconductivity [31] are a direct consequence of the interaction between electrons and lattice vibrations. The e-p interaction also plays an important role in the materials that can be utilized for thermoelectric devices [32,33]. Other fundamental physical phenomena such as the Kohn anomaly [34] and the Peierls [35] distortions are also direct consequences of the e-p interaction. The same is also responsible for the broadening of the spectral lines as observed in angle-resolved photoemission spectroscopy [36] and in vibrational spectroscopy [37], as well as for the temperature dependence of the band gaps in semiconductors [38]. The calculation of the el–ph coupling from first-principles is challenging because of the necessity of evaluating Brillouin zone integrals with a high accuracy. Such calculation requires the evaluation of matrix elements between electronic states connected by phonon wavevectors [39]. Well-established simulation packages are available for computing electronic states and eigenvalues using the DFT framework [26], and, the phonon frequencies and eigenmodes by utilizing density-functional perturbation theory (DFPT) [40]. However, this does not eliminate the need of large numbers of matrix elements that are necessary for achieving numerical convergence of the Brillouin zone integrals.

In present article, we have studied the carrier mobility of bulk MgSe and MgTe systems. The crucial parameter, e-p coupling has been included for precisely computing the carrier mobility within the materials. We have given more stress on the term "mobility", as it can shed light on the questions of central interest like, 1. How the carriers move inside a crystal? 2. What are the constrains on their motion? 3. What parameters influence their motion? The detailed methodology is described in Computational Details section, and the results are discussed in the Results & Discussion section. Extract of the article is summarized in the Conclusion section.

## 2. COMPUTATIONAL DETAILS

Accurate prediction of transport properties can set-up a benchmark in real sense as it serves as a stepping stone for designing materials with desired transport properties. The early journey of predicting carrier mobility, points towards the semi-empirical models that are better in predicting electron lifetimes for isotropic materials with simple parabolic bands. The deformation potential theory proposed by Bardeen and Shockley [41] is one of the wide-spread theoretical method that incorporates the modulation of the electronic edge states together with the applied deformation/strain and the elastic moduli of the material. This is also limited to some extent, as it only accounts the contributions from the acoustic phonon mode dependent phonon scattering and ignores the scattering/perturbations caused by transverse phonon modes and, the deformation caused under anisotropic and complex crystals. In second method, electron lifetimes are considered as constant for all electronic states. Combination of the same with the Fourier or Wannier interpolation of ab initio electronic band structure has been proven to be most efficient way for the treatment of complex anisotropic crystals having non-parabolic bands [27]. Final method that incorporates DFPT and Wannier interpolation (EPW) has been proven to incorporate computation of the e-p coupling matrix that yields fair and comparable results having remarkable agreement with the experimental measured data [42].

The calculation is performed based on DFT framework and implemented within the plane wave pseudopotential simulation package Quantum Espresso (QE) [43]. The electronic and core states of the Mg, Se and Te atoms are described using norm conserving pseudopotentials together with local density approximated exchange-correlation functional. The optimization done with basic DFT calculation also the phonon dispersion curve and band structure are compared between DFT calculation and EPW part of the calculation. The optimized magnitude of kinetic energy cut-off for wave function was found to be 90 Ry, with the $\Gamma$-centered Monkhorst-Pack [44] k-mesh grid with 16x16x16 density. The unit cells were completely relaxed with respect to all fundamental cell parameters and atomic positions until achieving the convergence of total energy with respect to the threshold parameters. The phonon dispersion curve was computed for understanding the lattice dynamics and for further computation of carrier mobility under EPW approach. The **q**-mesh of 8x8x8 was found sufficient for getting converged results.

As mentioned in previous section, the calculation of the el–ph coupling from *first-principles* is challenging because the evaluation of the Brillouin zone integrals demands high accuracy. Such calculation requires the evaluation of matrix elements between electronic states connected by phonon wave vectors [45]. Well-established software packages are available for computing electronic states and eigenvalues through DFT [43,46,47], as well as phonon frequencies and eigen modes through DFPT [48]. However, as a known fact, large number of matrix elements are critical for achieving numerical convergence of the Brillouin zone integrals. The carrier mobility is defined as,

$$\mu^{\alpha\beta} = \frac{\sigma^{\alpha\beta}}{n_c q}.$$

Where, $n_c$ is the density of the carriers. The scattering rates of electrons and holes can be calculated using Fermi golden rule accounting for a phonon bath in equilibrium at room temperature (T = 300 K) [49]:

$$\frac{1}{\tau_{i,\mathbf{k}}^{\nu}} = \frac{2\pi}{\hbar} \sum_{\mathbf{q},j} \left|g_{\mathbf{qk}}^{(ij)\nu}\right|^2 \left[N_{\nu,\mathbf{q}}\delta\big(E_{j,\mathbf{k}+\mathbf{q}} - \hbar\omega_{\nu,\mathbf{q}} - E_{i,\mathbf{k}}\big) + \big(N_{\nu,\mathbf{q}} + 1\big)\delta\big(E_{j,\mathbf{k}-\mathbf{q}} + \hbar\omega_{\nu,\mathbf{q}} - E_{i,\mathbf{k}}\big)\right]$$

$h$ is the reduced Planck's constant, $\varepsilon$ is the electron energy, $\delta$ is the Dirac delta function and g is the coupling matrix element.

## 3. RESULTS AND DISCUSSION

The optimized bulk crystal structure of the system is presented in the Figure 1(a). Akin to III-V materials [50], these materials also favour cubic zinc blende crystal structure in their bulk phase which is isotropic in nature with may be result in less complex electronic and phonon band dispersions. The optimized lattice constants of MgSe and MgTe are 5.92 Å and 6.39 Å, respectively which are in excellent agreement with the reported experimental values [51–53]. It is noteworthy that the precise computation of structural parameters symbolises the accurate prediction of the inter-atomic forces within the prescribed threshold criteria, which is a necessary pre-requisite for correctly predicting the ground state properties of the materials. As observed from Figures 1(b) and (c), both materials possess direct bandgap, that is highly desired for designing high-performance optoelectronic devices. As expected, the bandgap of MgSe (2.95 eV) is higher than the MgTe (2.59 eV). Further, the EPW incorporated electronic bandgaps were found to be 2.43 eV and 2.29 eV for MgSe and MgTe, respectively. These values are also supported by the reported data available in literature [54–56]. Most of the energy applications rely on the electronic band structure and edge state curvature of the material to be used; as the direct band gap accounts for the optical transitions, and the crisp band curvature is responsible for diverse carrier effective mass. The computed results advocate the utility of these materials in the field of photocatalyst/photovoltaic and optoelectronic devices owing to their unique band edge and direct electronic gaps.

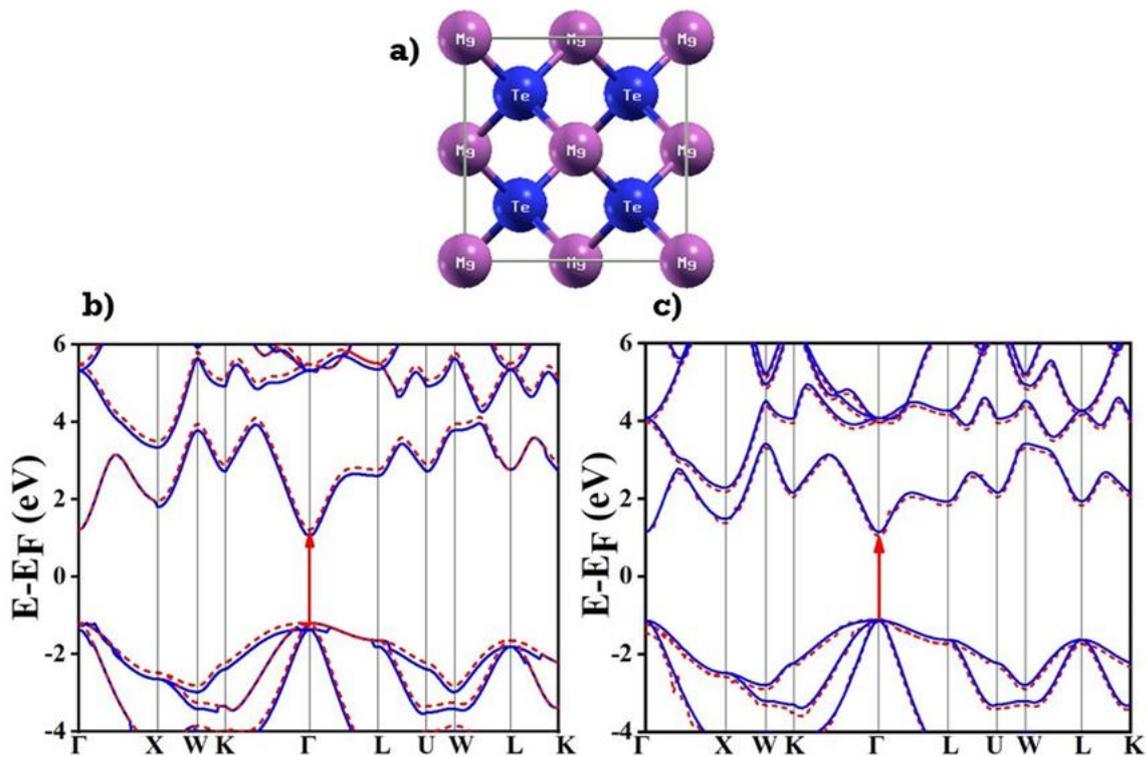

Figure 1. (a) Optimized crystal structure of MgSe/Te with cubic zinc blende symmetry. The atoms in blue represent chalcogen atoms (Se/Te), and the purple atoms are indicating Mg atoms, electronic band structure of (b) MgSe and (c) MgTe. The solid blue line and dashed red lines represent DFT and EPW computed band structures, respectively.

As expected, the EPW computed bandgap is less than the DFT computed bandgap (See Figs. 1(b) and (c)). This is because of the reason that on incorporating the e-p interaction, the contributions of lattice vibrations are accounted. This results in softening of the electronic dispersion, thereby reducing the gap between the conduction and valence band edges [29]. Post the electronic properties, the most important property i.e., phonon dispersion curves for MgSe and MgTe were computed (See Figs. 2(a-d)). All computed phonon frequencies are observed to be positive suggesting dynamical stability of the systems.

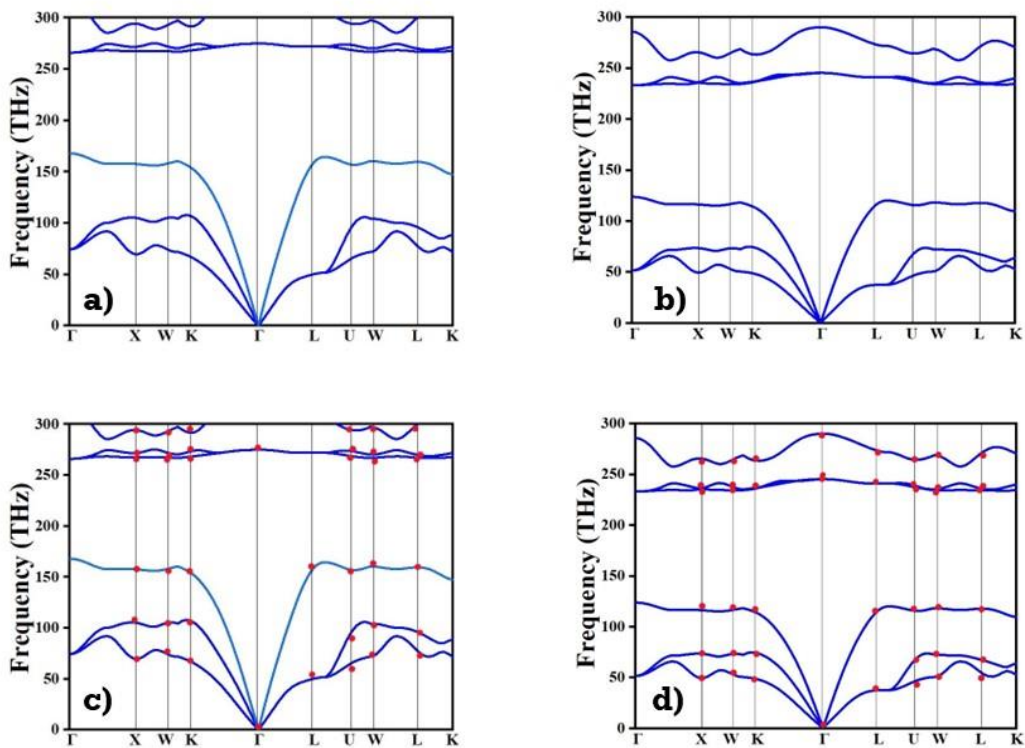

Figure 2. Computed phonon dispersion curves for MgSe (a) and (c), and MgTe (b) and (d), respectively. The (a) and (b) curves represent DFT computed phonon dispersions, and the (c) and (d) curves represent phonon dispersions post incorporating EPW approach.

Akin to the band structures, the EPW based phonon dispersion curves of MgSe and MgTe also show softening in the magnitudes. This causes significant reduction in the vibrational frequencies. This again is the result of the markable interaction between the electronic states and the vibrational modes that causes modulation in the system potential which is responsible for phonon renormalization. Furthermore, the modifications in the system temperature may cause pronounced renormalization effects which also is responsible for enhanced anharmonicity in the systems [42,57]. The total atomic contribution for both systems is two atoms per unit cells, thereby producing total 2X3 = 6 phonon modes. As usual, the three acoustic phonon modes possess zero frequency at the $\Gamma$-point; i.e., the centre of the Brillouin zone. The remaining three branches in the phonon dispersion represent optical phonon modes. These three optical modes can be further classified as one longitudinal and two transverse optic phonon modes, respectively. Understanding the characteristic signatures and trend of these modes provides key insights to the transport mechanism of the system. One such characteristic signature which is usually observed in most of the non-centrosymmetric crystals is the LO-TO splitting; this gets further enhanced for the systems that involve atoms with significant ionicity and possess strong polarization character. Furthermore, it is interesting to note that the e-p coupling also influences this LO-TO gap. Reduction in the LO-TO splitting can be the result of weaker e-p coupling, and vice-versa.

After investigating the phonon dispersion curve, we move towards the final objective of the present work. The evaluation of carrier mobility of the systems. For proper description of the carrier mobility, the e-p associated carrier scattering rates are computed. As observed from the Figure 3, the MgSe (a) and MgTe (b) both the materials indicate dense population across selected energy levels implicating that the respective associated carriers are strongly scattered by the phonons and vice-versa. Further, the computation of carrier mobility was done for both systems. It was observed that the carrier concentration of $10^{18}$ cm$^{-3}$ was found optimum for further computation of the temperature dependent mobility. As expected, the electron mobility (c) for both systems is found to be significantly higher than the hole mobility (d). Interestingly, the hole and electron mobilities of MgSe (Blue) and MgTe (Red) show different trends. For electron mobility, MgSe possess enhanced magnitude as compared to the MgTe (See Fig. 3(c)), in contrast to the hole mobility, which shows pronounced trend for MgTe as compared to the MgSe. This is in agreement with the observed carrier scattering rates (See Fig. 3(a) & (b)). The Fig. 3(a) indicates that the scattering rates post fermi level show lesser density as compared to the negative energy regime. This indicates that the electrons in MgSe experience weaker scattering as compared to the holes, thereby following the trend as shown in the Fig. 3(c) and (d). On the other hand, reverse trend in case of MgTe is observed (See Fig. 3(b)), thereby supporting the modified trend as observed for temperature dependent carrier mobility (See Fig. 3(c) and (d)).

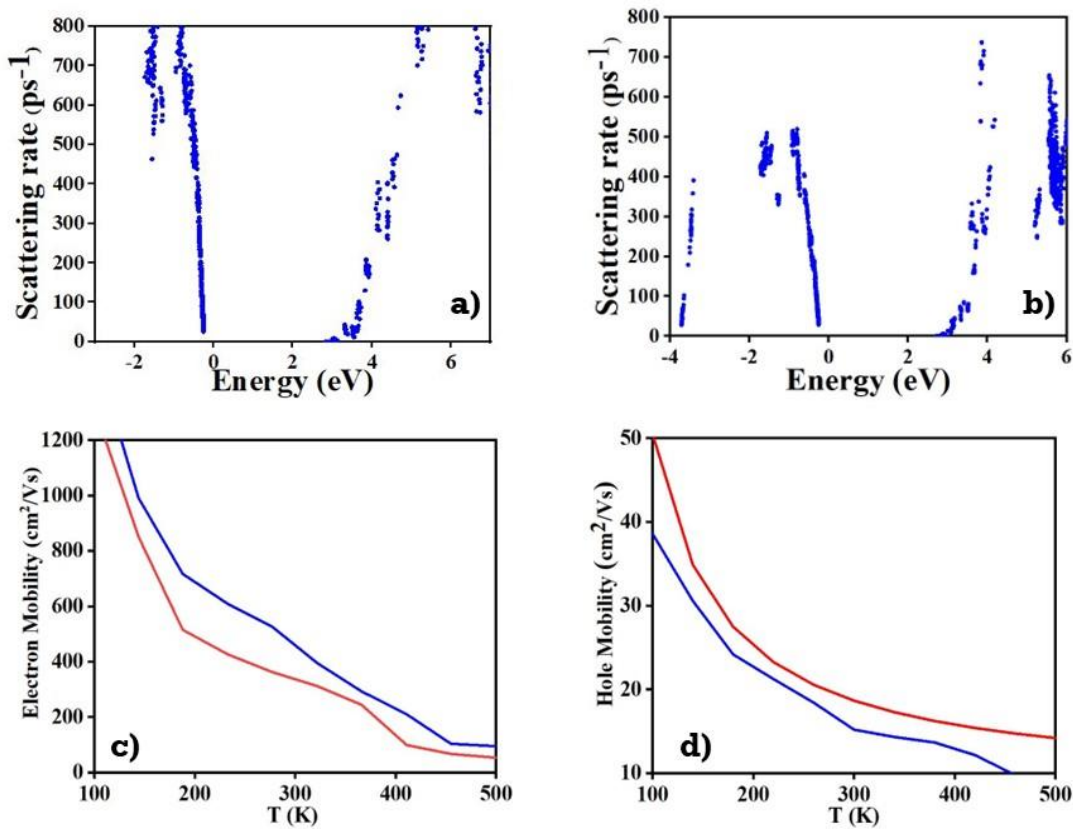

Figure 3. Computed scattering rate for (a) MgSe and (b) MgTe and, temperature dependent electron (c), and hole (d) mobility for MgSe (blue curve) and MgTe (red curve), respectively under EPW approach (carrier concentration: $10^{18}$ cm$^{-3}$).

Further, the computed room temperature electron and hole mobility for MgTe is 335 cm$^2$/Vs and 18.7 cm$^2$/Vs, respectively. These results can be compared with the results of two famous semiconductor materials, i.e., GaN [58] and Si [59]. In case of Si, the electron mobility is found to be <400 cm$^2$/Vs and, hole mobility as <50 cm$^2$/Vs at $10^{18}$ cm$^{-3}$ carrier concentration. This supports that the present materials can be further considered for further studies for testing their utility in the future devices. Furthermore, in case of CaTe, the hole mobility at 300K is approximately 20 cm$^2$/Vs [60]. In addition, the alloy systems, CdMgTe possess mobility of 453 cm$^2$/Vs [61] and, at carrier concentration of $10^{19}$ this mobility decreases to 190

cm$^2$/Vs [61]. Several other Te based alloys also have been reported to possess mobility in the range of ~ 100 cm$^2$/Vs [62,63].

## 4. CONCLUSION

The DFT based computation of the ground state electronic, vibrational and carrier dynamics of bulk MgSe and MgTe under cubic zinc blende phase was performed. For accurate description of the carrier dynamics, incorporation of EPW approach was done. It is observed that the EPW associated results are in excellent agreement with the previously reported data. This is subjected to the fact that the inclusion of e-p interaction correctly accounts for the contributions arising due to the e-p coupling. Comparing the pure DFT and EPW treated band and phonon dispersions of the systems show softening effect implicating the strength of the coupling between the e-p. Furthermore, the computed scattering rate of the carriers resembles the expected behaviour that can be justified by the electronic band structures and density of states trends. The targeted property of the systems, i.e., the carrier mobility associated with both the materials show similar trend as the reported materials that are being used for fabricating the electronic devices with significantly lower magnitude. It is believed that the present study sheds light on the fundamental aspects of the carrier dynamics within the MgSe and MgTe thereby suggesting that further improvement in the system properties can be done by means of dimensional confinement or by modulating the chemical composition of the systems.

## ACKNOWLEDGEMENT(S)

The authors acknowledge the support received from The Maharaja Sayajirao University of Baroda, Vadodara, Gujarat, India. One of the authors T.K.G. acknowledges Department of Science and Technology (DST), Ministry of Science and Technology, Government of India, New Delhi for providing financial support (Grant no.: SR/WOS-A/PM-95/2016 (C)).

## STATEMENT OF DECLARATION

No competing/non-competing interest(s) to declare.

## AUTHOR CONTRIBUTIONS

MJ: Computation, draft preparation; TKG & PKJ: Designing the problem, writing, editing & correcting the manuscript.

## DATA AVAILABLITY

The data can be made available on reasonable request to the corresponding author.

(2017).